\documentclass[pre,preprint,groupedaddress,showpacs]{revtex4}

\usepackage{amsmath}
\usepackage{amssymb}
\usepackage{amsfonts}
\usepackage{mathrsfs} 
\usepackage{epsfig}
\usepackage{bm}
\DeclareMathAlphabet{\mathpzc}{OT1}{pzc}{m}{it}

\begin{document}

\title{Ergodic model for the expansion of spherical nanoplasmas}

\author{F. Peano$^{1,2}$}\email{fabio.peano@ist.utl.pt}
\author{G. Coppa$^{1}$}
\author{F. Peinetti$^{1}$}
\author{R. Mulas$^{1}$}
\author{L. O. Silva$^{2}$}
\affiliation{$^1$Dipartimento di Energetica, Politecnico di Torino, 10129 Torino, Italy}\author{ }
\affiliation{$^2$GoLP/Centro de F\'isica dos Plasmas, Instituto Superior T\'ecnico, 1049-001 Lisboa, Portugal}

\date{\today}

\begin{abstract}
Recently, the collisionless expansion of spherical nanoplasmas has been analyzed with a new ergodic model, clarifying the transition from hydrodynamic-like to Coulomb-explosion regimes, and providing accurate laws for the relevant features of the phenomenon. A complete derivation of the model is here presented. The important issue of the self-consistent initial conditions is addressed by analyzing the initial charging transient due to the electron expansion, in the approximation of immobile ions. A comparison among different kinetic models for the expansion is presented, showing that the ergodic model provides a simplified description, which retains the essential information on the electron distribution, in particular, the energy spectrum. Results are presented  for a wide range of initial conditions (determined from a single dimensionless parameter), in excellent agreement with calculations from the exact Vlasov-Poisson theory, thus providing a complete and detailed characterization of all the stages  of the expansion.
\end{abstract}

\maketitle

\newcommand{\ped}[1]{_{\text{#1}}}\newcommand{\api}[1]{^{\text{#1}}}
\newcommand{\diff}[1]{\text{d}#1}

\section{Introduction}

The irradiation of solid targets with ultraintense lasers can induce the prompt formation of hot dense plasmas, which rapidly expand into a vacuum, as predicted by J. Dawson in 1964 \cite{Dawson}.
In the case of planar targets (employed for ion acceleration \cite{ion_acc}), the physics of the expansion has been extensively studied under a variety of conditions, using different analytical and numerical approaches \cite{planar,Crow,Mora,Betti,Manfredi}.
In contrast, the expansion of spherical plasmas [such as the nm- to $\mu$m-sized plasmas generated upon interaction of ultraintense lasers with atomic or molecular clusters (cf. \cite{Ditmire_Nature,fusion_exp,Sakabe,Last,Heidenreich})] have not been analyzed as thoroughly. A deep knowledge of the expansion (accounting for the self-consistent dynamics of ions and electrons) can be relevant in particular situations where accurate control over the expansion is necessary, examples being the double-pump irradiation of deuterium clusters aimed at tailoring the ion dynamics so as to induce intracluster fusion reactions \cite{Peano_PRA}, or the biomolecular imaging with ultrashort X-ray pulses \cite{xrays}, where expansion control is needed to avoid significant damages of the sample before the typical imaging time.
In fact, accurate solutions for spherical expansions exist only for ideal cases, such as the Coulomb explosion (CE) \cite{Kaplan_PRL} of a pure ion plasma, which occurs when all the electrons are suddenly swept away from the cluster by the laser field. In opposite conditions, when most of the electrons are heated by the laser but not stripped from the cluster, hydrodynamic models can be employed  \cite{hydrodynamic} to estimate the basic features of the expansion; in the quasineutral limit, a kinetic solution for the adiabatic expansion of plasma bunches into a vacuum has also been derived \cite{Kovalev}.
However, in more general situations (for example, in experiments with large clusters, containing millions of atoms), a significant violation of charge neutrality occurs, even though a relevant fraction of the electrons remains bound to the cluster. In such conditions (corresponding to plasma radii on the order of the electron Debye length, or less), the expansion process is strongly dependent on the self-consistent dynamics of ions and trapped electrons and it can be described accurately only by kinetic models, based on the Vlasov-Poisson (VP) theory. A remarkable example of numerical solution of the VP equations for the expansion problem in a spherical geometry can be found in Ref. \cite{Manfredi}, for the particular case in which the motion of both ions and electrons is purely radial.

Recently, the Authors presented a kinetic analysis of the collisionless expansion of spherical plasmas driven by hot electrons, based on a peculiar ergodic model, which accounts for the radial motion of the ions and for the three-dimensional motion of nonrelativistic electrons  \cite{Peano_PRL_2}. In the present paper, the model is derived in detail,  and its validity is tested against reference solutions of the full VP equations (here obtained using ad-hoc numerical techniques).
Furthermore, a procedure to determine the self-consistent initial conditions for the expansion within the framework of the model is presented.
As an accurate knowledge of the initial space-charge distribution is fundamental to describe correcly the long-term plasma expansion, the initial charging transient, during which the faster electrons leave the cluster core, is analyzed resorting to different models, thus providing deeper physical insights and validating the technique.

The results presented here provide a complete characterization of the expansion dynamics of spherical nanoplasmas, which can be useful in the interpretation of recent experiments with clusters, either irradiated with intense IR lasers or with VUV/X-ray sources \cite{xrays}, where conditions may be far from those of a pure CE \cite{Sakabe}. In particular, the different behavior observed in the ion energy spectrum for different values of the electron temperature allows a clear identification of the transition from hydrodynamic-like to CE expansion regimes, thus defining the range of validity of the CE approximation. 

\section{Kinetic models for the expansion}
\label{sec:model}
In the electrostatic, nonrelativistic limit, the dynamics of a collisionless ion-electron plasma is described rigorously by the Vlasov-Poisson (VP) set of equations:
\begin{equation}
\left\{\begin{array}{rcl}
	\dfrac{\partial f\ped{e}}{\partial t}	&=& -{\bf v}\cdot\dfrac{\partial f\ped{e}}{\partial {\bf r}} -\dfrac{e}{m}\dfrac{\partial \Phi}{\partial {\bf r}}\cdot\dfrac{\partial f\ped{e}}{\partial {\bf v}} \vspace{0.25cm}\\
	\dfrac{\partial f\ped{i}}{\partial t}	&=& -{\bf v}\cdot\dfrac{\partial f\ped{i}}{\partial {\bf r}} +\dfrac{Ze}{M}\dfrac{\partial \Phi}{\partial {\bf r}}\cdot\dfrac{\partial f\ped{i}}{\partial {\bf v}} \vspace{0.15cm}\\
	\nabla^2\Phi &=& \displaystyle 4\pi e\left(\int f\ped{e}\diff{{\bf v}} -Z\int f\ped{i}\diff{{\bf  v}} \right)
\end{array}\right. \text{,}
\label{eq:VP_1}
\end{equation}
where $f\ped{e}({\bf r},{\bf v},t)$ and $f\ped{i}({\bf r},{\bf v},t)$ are the distribution functions in phase space for electrons (having mass $m$ and charge $-e$) and ions (having mass $M$ and charge $Ze$), respectively, and $\Phi\left({\bf r},t\right)$ is the electrostatic potential ($\Phi$ is set to zero at infinity, so that the energy of a single electron, $\epsilon=\frac{1}{2}mv^2-e\Phi$, is negative if it is trapped). 
In the following, the attention is focused on the expansion of a plasma sphere (with initial radius $R_0$) composed of cold ions (with initial uniform density $n\ped{i0}$) and hot electrons [with initial uniform density $n\ped{e0}=Zn\ped{i0}$ and arbitrary energy distribution $\rho\ped{e}(\epsilon)$]. The general initial conditions for Eqs. \eqref{eq:VP_1} can be cast in the form
\begin{equation}
\left\{\begin{array}{rcl}
	f\ped{e0}\left({\bf r},{\bf v}\right) &=& n\ped{e0}g\left(v\right)\Theta\left(1-\dfrac{r}{R_0}\right) \vspace{0.15cm}\\
	f\ped{i0}\left({\bf r},{\bf v}\right) &=& n\ped{i0} \delta \left({\bf v}\right)\Theta\left(1-\dfrac{r}{R_0}\right)
\end{array}\right.\text{,}
\label{eq:f_t0}
\end{equation}
where $\Theta$ is the Heaviside step function and $g$ is an arbitrary function of $v,$ such that $\rho\ped{e}(\epsilon)$ $=$ $16\pi^2n\ped{e0}R_0^3/3$ $(2\epsilon/m^3)^{1/2}$ $g[(2\epsilon/m)^{1/2}]$.
In the present paper, for simplicity, only the ideal situation (commonly adopted in the literature \cite{planar,Crow,Mora,Betti,Manfredi}) of an initially neutral plasma with Maxwellian electrons is considered, in which all the information on the electron heating by the laser pulse is contained in the initial electron temperature, $T_0$. The function $g$ in Eq. \eqref{eq:f_t0} is then defined as
$g\left(v\right) =  \left[m/(2\pi k\ped{B}T_0)\right]^{3/2}\exp\left[-mv^2/(2k\ped{B}T_0)\right]$.
As can be readily proved by writing Eqs. \eqref{eq:VP_1} and \eqref{eq:f_t0} in nondimensional form, in this case the dynamics of the system is fully determined by the dimensionless parameters $Zm/M$ and $\widehat{T}_0 = Zk\ped{B}T_0/\epsilon\ped{CE}=3\lambda\ped{D0}^2/R_0^2$, being $\lambda\ped{D0}$ the initial Debye length for the electrons, and $\epsilon\ped{CE}=\frac{4\pi}{3}e^2 R_0^2 n\ped{i0}^2$ the maximum ion energy attainable in the case of pure CE for a sphere of ions.
With the use of the initial conditions \eqref{eq:f_t0} the electrons are supposed to be instantaneously heated by an infinitely short laser pulse, without expanding. However, in principle, any initial space-energy density for the electrons could be employed (for example, linear superpositions of Maxwellian distributions, as well as initially nonneutral distributions), as resulting from a realistic model of laser-matter interaction. Nonetheless, the use of the reference initial conditions \eqref{eq:f_t0} provides a simple way to obtain useful physical insights on the influence of the combined effects of variations of electron energy and cluster features. 

The expansion process is split in two stages: first, a rapid expansion of the electrons, which leads to a VP equilibrium before the ions move appreciably; second, a slow expansion of the plasma bulk, driven by the positive charge buildup formed in the first stage. Due to the large mass disparity between ions and electrons, a simplified model can be derived, in which the ions are assumed as immobile during the former process, whereas the electrons can be considered as instantaneously at equilibrium with the electrostatic potential during the latter stage. A self-consistent theoretical framework can be developed, which allows one to determine accurately both the initial equilibrium and the bulk expansion, by treating the electron dynamics as a sequence of equilibrium configurations (with frozen ions). The model is obtained by exploiting the functional relation 
existing between $n\ped{e}$ and $\Phi$ at equilibrium, and by calculating the energy variation of the electrons under the hypothesis of slow variations of $\Phi$ in time.

Equilibrium solutions of the Vlasov equation for the electrons must depend on ${\bf r}$ and ${\bf v}$ only through the invariants of motion. Since a spherical symmetry has been assumed and the electrostatic force is central, the only invariants of motion to be considered are the Hamiltonian $\mathscr{H}\left({\bf r},{\bf v}\right) = m v^2/2-e\Phi\left({r}\right)$ and the angular momentum, ${\bf L}= m{\bf r}\times{\bf v}$. Consequently, the equilibrium distribution function can be written as $f\ped{e}\left({\bf r},{\bf v}\right) = F\left(\mathscr{H}\left({\bf r},{\bf v}\right),m{\bf r}\times{\bf v}\right)$. 
If a generic space point ${\bf r}=r\hat{\bf e}_r$ and a generic velocity ${\bf v}=v_r\hat{\bf e}_r+v_{\bot}\hat{\bf e}_{\bot}$ are considered, the phase-space density is given by $F\left(\dfrac{m}{2}\left(v_r^2+v_{\bot}^2\right)-e\Phi\left(r\right), \mathscr{L},\hat{\bf e}_r\times\hat{\bf e}_{\bot}\right)$ being $\mathscr{L}=mrv_{\bot}$ the absolute value of ${\bf L}$. Due to the spherical symmetry of the system (and, in particular, the symmetry with respect to any rotation with respect to $\hat{\bf e}_r$), the phase-space density cannot depend upon $\hat{\bf e}_{\bot}$, and, consequently, $F$ depends only on $\mathscr{H}$ and $\mathscr{L}$.

The energy-angular momentum distribution, $\sigma\ped{e}\left(\epsilon,\ell\right)$, can be defined as
\begin{align}
	\sigma\ped{e}\left(\epsilon,\ell\right) =& \iint F\left[\mathscr{H}\left({\bf r},{\bf v}\right),\mathscr{L}\left({\bf r},{\bf v}\right)\right]\delta\left[\mathscr{H}\left({\bf r},{\bf v}\right)-\epsilon\right]
\delta\left[\mathscr{L}\left({\bf r},{\bf v}\right)-\ell\right]\diff{{\bf r}} \diff{{\bf v}} = \nonumber\\
& \dfrac{8\pi^2\sqrt{2}}{m^{3/2}} f(\epsilon,\ell)\displaystyle\int_{R_1\left(\epsilon,\ell\right)}^{R_2\left(\epsilon,\ell\right)} \left[ \epsilon - \dfrac{\ell^2}{2mr^2} +e\Phi(r)\right]^{-\frac{1}{2}} \diff{r} \text{,}
\label{eq:rho_e_l}
\end{align}
where $R_1\left(\epsilon,\ell\right)$ and $R_2\left(\epsilon,\ell\right)$ $\left(R_1\le R_2\right)$ are the radial turning points, i.e., the values of $r$ such that $\epsilon-\frac{\ell^2}{2mr^2}+e\Phi\left(r\right)=0$.
The quantity $\sigma\ped{e}\left(\epsilon,\ell\right)\Delta\epsilon\Delta\ell$ represents the number of electrons having energy in $[\epsilon,\epsilon+\Delta\epsilon]$ and absolute value of the angular momentum in $[\ell,\ell+\Delta\ell]$. The electron density, $n\ped{e}$, can be written as
\begin{equation}
	n\ped{e}(r) = \dfrac{1}{4\pi r^2}\iint \sigma\ped{e}\left(\epsilon,\ell\right) \mathcal{P}\left(r,\epsilon,\ell;\{\Phi\}\right)\diff{\epsilon}\diff{\ell} \text{,}
\label{eq:n_e_l}
\end{equation}
where
\begin{equation}
	\mathcal{P}\left(r,\epsilon,\ell;\{\Phi\}\right) = \dfrac{ \left[ \epsilon - \dfrac{\ell^2}{2mr^2} +e\Phi\left(r\right)\right]^{-\frac{1}{2}} }{ \displaystyle\int_{R_1\left(\epsilon,\ell\right)}^{R_2\left(\epsilon,\ell\right)} \left[ \epsilon - \dfrac{\ell^2}{2m{r^{\prime}}^2} +e\Phi\left(r^{\prime}\right)\right]^{-\frac{1}{2}}\diff {r^{\prime}}} 
\label{eq:pr_e_l}
\end{equation}
is such that $\mathcal{P}(r;\epsilon,\ell)\Delta r$ gives the probability, for an electron with energy $\epsilon$ and angular momentum $\ell$, to be found in $[r,r+\Delta r]$. If time variations of $\Phi$, due to the ion motion, are slow with respect to the period of the of the radial oscillation of the electrons, the mean value of $\diff{\epsilon}/\diff{t}$ can be evaluated as
\begin{equation}
\Big\langle\dfrac{\diff{\epsilon}}{\diff{t}}\Big\rangle=\Big\langle-e\dfrac{\partial\Phi}{\partial t}\left(r\left(t\right),t\right)\Big\rangle=-e \displaystyle\int_{R_1\left(\epsilon,\ell\right)}^{R_2\left(\epsilon,\ell\right)} \dfrac{\partial \Phi}{\partial t}\left(r,t\right)\mathcal{P}\left(r,\epsilon,\ell;\right\{\Phi\left\}\right)\diff{t}
\label{eq:de_dt_e_l} 
\end{equation}
i.e., by using the ensemble average of $\dfrac{\partial \Phi}{\partial t}$. This is equivalent to preserve the value of the adiabatic invariant \cite{Goldstein}
\begin{equation}
\mathcal{I}\left(\epsilon\left(t\right),\ell,t\right)=\displaystyle\oint p_r\diff{r}=\textrm{Const}\cdot\displaystyle\int_{R_1\left(\epsilon,\ell\right)}^{R_2\left(\epsilon,\ell\right)}\left[\epsilon-\dfrac{\ell^2}{2mr^2}+e\Phi\left(r,t\right)\right]^{\frac{1}{2}}dr
\label{eq:inv_adi}
\end{equation}
Equations \eqref{eq:rho_e_l}-\eqref{eq:de_dt_e_l}, coupled with Poisson's equation and Newton's equation for the radial motion of the cold ions, provide a self-consistent model for the collisionless expansion of a finite-size plasma in the case of spherical symmetry. 

In kinetic theory, there is a precise relationship between time scales and the proper number of parameters to be used to describe correctly a given phenomenon: in the case of the plasma expansion, the VP system \eqref{eq:VP_1}  for $f\ped{e}$ and $f\ped{i}$ allows one to follow precisely the expansion dynamics on the time scale of the fastest particles; to study the ion expansion, a quasi-equilibrium model, Eqs. \eqref{eq:rho_e_l}-\eqref{eq:de_dt_e_l}, can be used, in which the stationary solution of the Vlasov equation for the electrons is employed. In fact, as the Vlasov model is noncollisional, it does not contain a physical mechanism leading towards the equilibrium (the equations are time-reversible). To justify the use of the equilibrium distribution $f\left({\bf r},{\bf v}\right)=f\left[\mathscr{H}\left({\bf r},{\bf v}\right), \mathscr{L}\left({\bf r},{\bf v}\right)\right]$, one must suppose that the stationary solution of Vlasov equation is a good representation of the real electron distribution, once high-frequency fluctuations are eliminated; formally, this can be performed by introducing a dissipation mechanism (i.e., a suitable collision term into the Vlasov equation). In general, an approximate kinetic model can be regarded  as the result of introducing a particular collision term. For example, by using a binary collision term with sufficiently high collision frequency, $f\ped{e}$ tends towards the Maxwell-Boltzmann distribution [i.e., $f\left({\bf r},{\bf v}\right)=\textrm{Const}\cdot\exp\left(-\mathscr{H}\left({\bf r},{\bf v}\right)/k\ped{B}T\right)$], in which all the information is restricted to the temperature; in this case, a hydrodynamic description is obtained, whose domain of validity is confined to situations where $\widehat{T}_0\ll 1$. For larger values of $\widehat{T}_0$, the use of a proper energy spectrum is fundamental; in fact, the energy distribution presents a cutoff for $\epsilon=0$ (for $\epsilon>0$ the electrons are not confined and their stationary density must vanish) and this fact is hardly compatible with a Maxwellian distribution having a non-negligible fraction of electrons with $\epsilon>0$.

Within this framework, the approach of Ref. \cite{Peano_PRL_2} can be introduced by considering a collision term of the form
\begin{equation}
J\left(f\ped{e}\right)=-\nu \left(f\ped{e}-\bar{f}\ped{e}\right),\qquad \bar{f}\ped{e}=\dfrac{1}{4\pi}\oint f\ped{e}\left({\bf r},v\widehat{\bf \Omega},t\right)\diff\widehat{\bf \Omega}
\label{eq:coll_int}
\end{equation}
where $\widehat{\bf \Omega}$ is a unit vector and $\nu$ represents the collision frequency; the specific value of $\nu$ is irrelevant, as long as $1/\nu$ is much smaller than the characteristic time of the ion expansion.
In this case, the kinetic Vlasov equation for the electrons is replaced by the collisional equation 
\begin{align}
	\dfrac{\partial f\ped{e}}{\partial t} = & -{\bf v}\cdot\dfrac{\partial f\ped{e}}{\partial {\bf r}} -\dfrac{e}{m}\dfrac{\partial \Phi}{\partial {\bf r}}\cdot\dfrac{\partial f\ped{e}}{\partial {\bf v}}\nonumber\\
&-\nu f +\dfrac{\nu}{4\pi}\oint f\left( {\bf r}, v\widehat{\bf\Omega},t \right)\diff{\widehat{\bf\Omega}}\text{,}
\label{eq:Boltzmann}
\end{align}
where $\nu$ is the collision frequency and $\widehat{\bf\Omega}$ is a unit vector.
The collisions do not alter the electron energy, but change randomly their direction, driving $f\ped{e}$ towards an equilibrium distribution having the form $f\ped{e}\left({\bf r},{\bf v}\right)=f\left[\mathscr{H}\left({\bf r},{\bf v}\right)\right]$, a sort of ergodic density such that each electron has an equal probability to be found in every point of the hypersurface of phase-space having equation $\mathscr{H}\left({\bf r},{\bf v}\right)=\epsilon$. This is different from the usual ergodic distribution of the statistical mechanics, in which the state of a system of $N$ particles can be found with equal probability on the hypersurface $\mathscr{H}\left({\bf r}_1,{\bf r}_2,...,{\bf r}\ped{N};{\bf v}_1,{\bf v}_2,...,{\bf v}\ped{N}\right)=$ Const of the complete, $6N$-dimensional phase-space. In fact, the time derivative of the electron entropy $S\ped{e}=-\iint f\ped{e}\log\left(f\ped{e}\right)\diff{\bf r}\diff{\bf v}$ can be written in the form
\begin{equation}
\dfrac{\diff{S\ped{e}}}{\diff{t}}=\nu\iint\log\left(\dfrac{f\ped{e}}{\bar{f}\ped{e}}\right)\cdot\left(f\ped{e}-\bar{f}\ped{e}\right)\diff{\bf r}\diff{\bf v}\text{,}
\label{eq:entropy}
\end{equation}
which is always nonnegative unless $f\ped{e}=\bar{f}\ped{e}$. Therefore, a necessary condition for the distribution to be stationary is that $f\ped{e}$ must not depend on $\widehat{\bf \Omega}$ . Finally, the equilibrium distribution is a function of $\mathscr{H}$ and $\mathscr{L}$ that does not depend on $\widehat{\bf \Omega}$, and, consequently, it is a function of $\mathscr{H}$ only.

In the following, the approach will be referred as single-particle ergodic (SPE) method.
According to this approach, the equilibrium distribution function can be written simply as $f\ped{e}\left({\bf r},{\bf v}\right) = f\left[\mathscr{H}\left({\bf r},{\bf v}\right)\right]$, the dependence on $\mathscr{L}$ being lost, and Eqs. \eqref{eq:rho_e_l}-\eqref{eq:pr_e_l} are replaced by
\begin{align}
	\rho\ped{e}\left(\epsilon\right) =& \iint f\left[\mathscr{H}\left({\bf r},{\bf v}\right)\right]\delta\left[\mathscr{H}\left({\bf r},{\bf v}\right)-\epsilon\right]\diff{{\bf r}} \diff{{\bf v}} = \nonumber\\
& \dfrac{16\pi^2\sqrt{2}}{m^{3/2}} f(\epsilon)\displaystyle\int_{\mathscr{D}\left(\epsilon\right)}\left[ \epsilon +e\Phi(r)\right]^{\frac{1}{2}} r^2 \diff{r} \text{,}
\label{eq:rho_e}
\end{align}
\begin{equation}
	n\ped{e}(r) = \dfrac{1}{4\pi r^2}\int \rho\ped{e}\left(\epsilon\right) \mathcal{Q}\left(r,\epsilon;\{\Phi\}\right)\diff{\epsilon} \text{,}
\label{eq:n_e}
\end{equation}
\begin{equation}
	\mathcal{Q}\left(r,\epsilon;\{\Phi\}\right) = \dfrac{ r^2\left[ \epsilon+e\Phi(r)\right]^{\frac{1}{2}} }{ \displaystyle\int_{\mathscr{D}\left(\epsilon\right)} {r^{\prime}}^2\left[ \epsilon+e\Phi\left(r^{\prime}\right)\right]^{\frac{1}{2}}\diff{r^{\prime}}} \text{,}
\label{eq:pr_e}
\end{equation}
where $\mathscr{D}\left(\epsilon\right)$ is the integration domain, such that $r\in\mathscr{D}\left(\epsilon\right)\Rightarrow\epsilon+e\Phi\left(r\right)\ge0$ \{for monotonic potentials, $\mathscr{D}\left(\epsilon\right)=\left[0, R\left(\epsilon\right)\right]$, where $\epsilon+e\Phi\left(R\left(\epsilon\right)\right)=0$\}, and the adiabatic invariant, Eq \eqref{eq:inv_adi}, is replaced by the ergodic invariant \cite{Ott} 
\begin{equation}
	\mathcal{J}\left[\epsilon\left(t\right),t\right] = \textrm{Const}\cdot \int_{\mathscr{D}\left(\epsilon\right)} \left[\epsilon+e\Phi(r,t)\right]^{\frac{3}{2}} r^2 \diff{r} \text{,}
\label{eq:I_e}
\end{equation}
defined as the volume of the region of $\mathbb{R}^6$ enclosed by the hypersurface of equation $\frac{1}{2}mv^2-e\Phi(r) = \epsilon$. 

As shown in Ref. \cite{Peano_PRL_2}, the SPE approach provides excellent results for the expansion of a spherical plasma in a wide range of the parameter $\widehat{T}_0$. There is a number of reasons to explain its success, even though it is not easy to quantify their relative importance. First, even though hydrodynamic models can provide a qualitative agreement with the real expansion dynamics, the SPE model is extremely more flexible in describing the energy distribution of the electrons. In addition, in the cases considered here, the initial phase-space distribution, Eq. \eqref{eq:f_t0}, is assumed to be an SPE function. Moreover, it must be noticed that the angular momentum is invariant only in the case of perfect spherical symmetry. In practical situations, perturbations to that symmetry (e.g., an initial shape which is not perfectly spherical, or collisions with heavy particles) would cause a mixing in $\mathscr{L}$ distribution, and their effect could be taken into account by introducing a collision term such as the one in Eq. \eqref{eq:coll_int}.   

\section{Single-particle ergodic model}
\label{sec:SPE}

Under the hypothesis of SPE distribution, a self-consistent model for the expansion of a spherical plasma can be formulated, starting from Eqs.  \eqref{eq:rho_e}-\eqref{eq:I_e}, as follows \cite{Peano_PRL_2}. A Lagrangian approach can be used both for the ions (which move in radial direction, starting from the initial position $r_0$, with zero velocity) and for the electrons (whose energy $\epsilon$ evolves in time starting from the initial value $\epsilon_0$), by determining the ions trajectories $r\ped{i}\left(r_0,t\right)$, the electron energies $\epsilon\left(\epsilon_0,t\right)$, the ion density $n\ped{i}\left(r,t\right)$, the electron density $n\ped{e}\left(r,t\right)$, the electron energy distributions  $\rho\ped{e}\left(\epsilon,t\right)$, and the potential $\Phi\left(r,t\right)$ according to the set of equations
\begin{equation}
\left\{\begin{array}{l}
M\dfrac{\partial^2 r\ped{i}}{\partial t^2}=-Ze\dfrac{\partial \Phi}{\partial r}\left(r\ped{i}\right)\text{,}\vspace{.1cm}\\

\dfrac{1}{r^2}\dfrac{\partial}{\partial r}\left(r^2\dfrac{\partial \Phi}{\partial r}\right)=
4\pi e\left(n\ped{e}-Zn\ped{i}\right)\text{,}\\

n\ped{i}\left(r\ped{i}\right)=n\ped{i,0}\left(r_0\right)\dfrac{\left(r_0/r\ped{i}\right)^2}{\partial r\ped{i}/\partial r_0}\text{,}\\

n\ped{e}=\displaystyle\int \rho\ped{e}\left(\epsilon\right)\mathcal{Q}\left(r,\epsilon,\left\{\Phi\right\}\right)\diff{\epsilon}\text{,}\vspace{.1cm}\\

\rho\ped{e}\left(\epsilon,t\right)=\dfrac{\rho\ped{e,0}\left(\epsilon_0\right)}{\partial\epsilon/\partial\epsilon_0}\textrm{,}\vspace{.2cm}\\

\dfrac{\diff{}}{\diff{t}}\mathcal{J}\left(\epsilon\left(\epsilon_0,t\right),t\right)=0\text{,}
\label{eq:model}
\end{array}\right.
\end{equation}
where the evolution equations for the radial coordinates of the ions and the electron energies are coupled via the Poisson's equation.
The expansion dynamics is determined once the initial ion density $n\ped{i,0}$ and the electron energy distribution $\rho\ped{e,0}$ are given. In Eqs. \eqref{eq:model}, the electron density $n\ped{e}$ is expressed as the sum of the number of electrons having energy in $\left[\epsilon,\epsilon+\diff{\epsilon}\right]$ [i.e., $\rho\ped{e}\left(\epsilon\right)\diff{\epsilon}=\rho\ped{e,0}\left(\epsilon\ped{0})\diff{\epsilon\ped{0}}\right)$], multiplied by the probability for an electron with energy $\epsilon$ to be found at the radius $r$, according to the ergodic distribution. For simplicity, the ion density $n\ped{i}$ is written under the hypothesis of no ion overtaking $\left(\partial r\ped{i}/\partial r\ped{0}\ne 0\right)$ \cite{Peano_PhD}; however, the model can be easily generalized to include many-branched shock shells \cite{Kaplan_PRL,Peano,Peano_PhD,Peano_PRA} and different ion species.

The set of equations \eqref{eq:model} describes the expansion dynamics on the ion time scale; therefore, its numerical solution is much faster than solutions of the full VP model (where the electron time scale must be followed).     
The model is solved by calculating the radial trajectories of a set of representative ions and the energy variations of a set of computational particles. Each computational particle represents a given number of electrons, whose radial distribution is given by Eq. \eqref{eq:pr_e}. This description of the energy dependence corresponds to a suitable discretization of the integral in Eq. \eqref{eq:n_e} (which is similar to the description of the spatial dependence commonly adopted in particle-in-cell approach \cite{Birdsall}).

\section{Charging transient}
\label{sec:transient}
Since the initial configuration considered here [cf., Eqs. \eqref{eq:f_t0}] is far from equilibrium, the proper $\rho\ped{e,0}$, to be used in Eqs. \eqref{eq:model}, must be determined as the equilibrium configuration following the initial charging transient. 
Apparently, the SPE method cannot be of help to this purpose, since it is valid only for sufficiently smooth variations of $\Phi$ in time, a condition which is not met in the early stage, when the hot electrons are suddenly allowed to expand (as if a rigid wall, initially confining them, were instantaneously brought to infinity).
However, a procedure has been envisaged, which makes these equations suitable also for the analysis of the initial electron equilibrium, thus allowing the study of the whole expansion process (initial charging transient and bulk expansion) within the same theoretical framework.
Before describing this procedure, the initial equilibrium is analyzed in detail, accounting for the full electron dynamics (VP model).

\subsection{Reference solutions of the collisionless and collisional models}
\label{sec:Vlasov_Boltzmann}
Due to the importance of an accurate knowledge of the initial equilibrium configuration of the electrons for a correct analysis of the plasma expansion, reference results for the transient leading to the initial space-charge distribution of the plasma has been determined by solving Eqs. \eqref{eq:VP_1} numerically, in the hypothesis of immobile ions. In the numerical scheme adopted, computational particles representative of a given number of electrons are moved in space, under the action of the sum of the self-consistent electric field and the electric field due to the ion distribution. By resorting to the spherical symmetry of the system, the field generated by the electrons is evaluated using Gauss' law, as if each particle were actually a spherical shell, thus avoiding the use of a computational grid for solving Poisson's equation, and allowing for an infinite radial domain (similar techniques have been used to investigate the VP dynamics of one-dimensional plasmas \cite{sheet}; here, the validity of the method has been checked through comparisons with reference results from the 3D particle-in-cell code OSIRIS \cite{OSIRIS}).
The same framework has been used also to investigate numerically the effect of the presence of the collision term, Eq. \eqref{eq:Boltzmann}, which forces the system towards a SPE distribution. Such perturbations are introduced in the model by scattering randomly the computational particles, without changing their energy, according to the collision frequency $\nu$. 

\begin{figure}[!htb]
\centering \epsfig{file=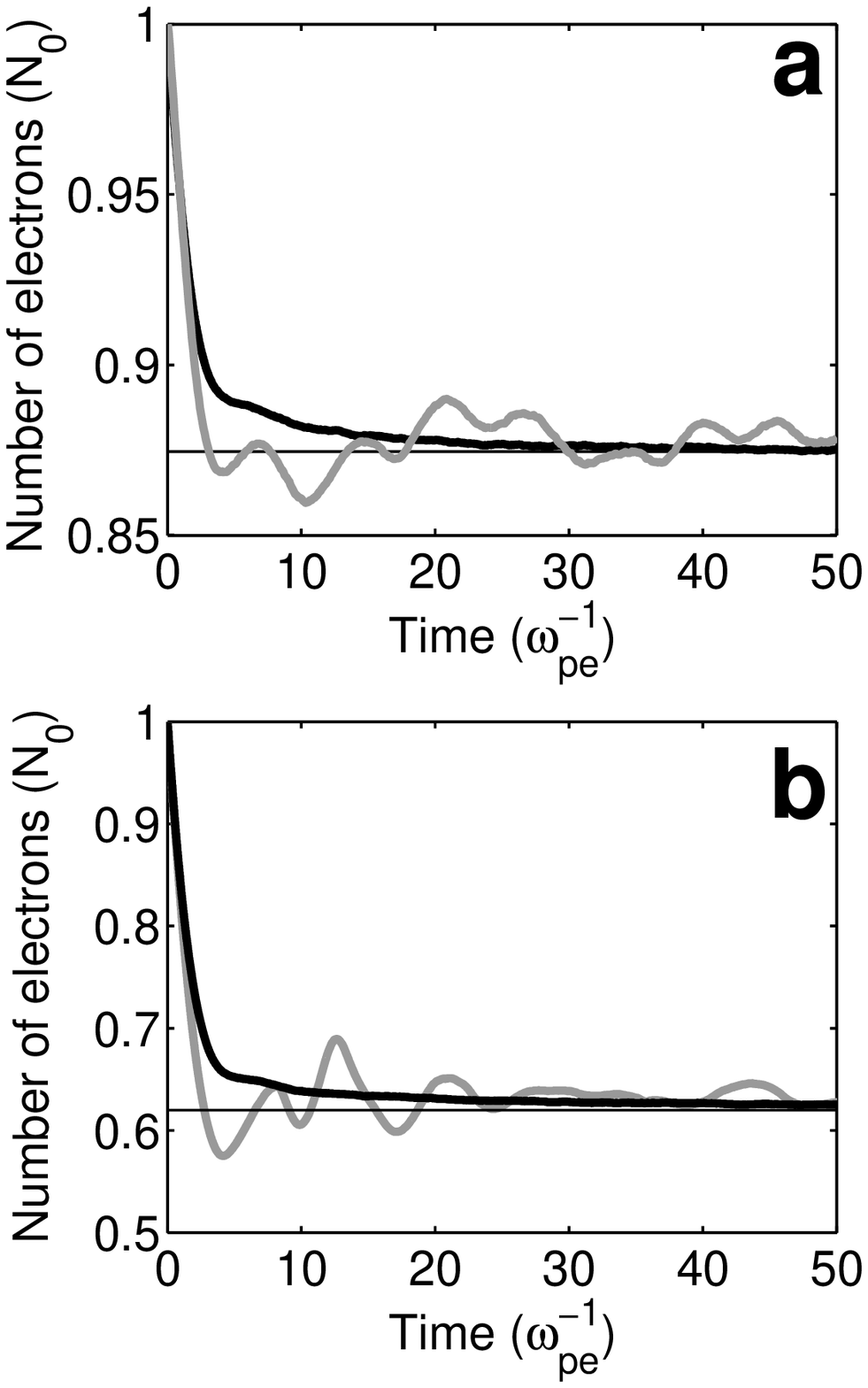, width=2.5in}
\caption{Evolution of the electronic charge contained within the ion sphere ($r<R_0$), for (a) $\widehat{T}_0 = 7.2\times 10^{-3}$ and (b) $\widehat{T}_0 = 7.2\times 10^{-2}$. Thick gray lines refer to the collisionless case, thick black lines refer to the collisional case ($\nu \sim \omega\ped{pe}$). Thin horizontal lines indicate the results obtained using the barrier method described in Sec. \ref{sec:barrier}. Units are normalized to the quantities indicated in parentheses.}
\label{fig:ch_tr_all}
\end{figure}

Figure \ref{fig:ch_tr_all} shows the evolution of the electronic charge contained within the ion sphere, for the representative low-temperature ($\widehat{T}_0=7.2\times10^{-3}$) and high-temperature ($\widehat{T}_0=7.2\times10^{-2}$) cases of Ref. \cite{Peano_PRL_2}, as obtained with the pure VP model ($\nu=0$) and with the collisional model, Eq. \eqref{eq:Boltzmann}, (using $\nu=\omega\ped{pe}$). In the collisionless case, the charge transient exhibits small-amplitude oscillations (the simulation parameters have been carefully checked to ensure that the oscillations are not due to numerical noise).
In the collisional model, for $\nu \gtrsim \omega\ped{pe}$, the oscillations are strongly damped and the system rapidly reaches an equilibrium configuration, as predicted theoretically.

\subsection{Method of the barrier}
\label{sec:barrier}
In order to build a self-consistent ergodic model for the whole expansion process (thus avoiding the use of different models to deal with the initial stage), a procedure has been devised to determine the equilibrium distribution that follows the initial electron expansion, using the same theoretical framework of Eqs. \eqref{eq:rho_e}-\eqref{eq:I_e}.
To this purpose, the charging transient described by the full VP model is replaced by a virtual charging transient, in which an external potential barrier, initially confining the electrons, is gradually moved from $R\ped{b}=R_0$ to infinity with a series of small radial displacements. Each time the barrier is moved farther by $\delta R\ped{b}$, the new self-consistent potential $\Phi$ is calculated and the energy of the electrons is updated. In order to actually simulate an expansion into vacuum (which the real transient is), the electrons and the expanding barrier must not exchange energy, i.e., the electron energy must vary only because of $\Phi$ variations. This implies that the ergodic invariant \eqref{eq:I_e} is not conserved during the initial stage. In fact, should one conserve $\mathcal{J}$ when displacing the barrier from a given radius $R\ped{b}$ to $R\ped{b} + \delta R\ped{b}$, the corresponding electron energy variation, $\delta\epsilon$, would be
\begin{equation}
	\delta\epsilon = -e\int^{R\ped{b}}_{0} \delta\Phi \mathcal{Q}\left(r,\epsilon;\{\Phi\}\right)\diff{r} - \delta W, \\
	\label{eq:delta_e_I_1}
\end{equation}
where $\delta W$, defined as
\begin{equation}
	\delta W = \frac{2}{3}\left[\epsilon+e\Phi\left(R\ped{b}\right)\right]
	\mathcal{Q}\left(R\ped{b};\epsilon\right)\delta R\ped{b}
	\text{.}
\label{eq:delta_e_I_2}
\end{equation}
represents the expansion work, done by an electron having energy $\epsilon$, against the expanding barrier. Thus, conserving $\mathcal{J}$ would cause the overestimation of the electron cooling as the system would lose an extra amount of energy corresponding to the expansion work. 
In order to obtain an energy balance equivalent to that of a vacuum expansion, the energy loss associated to the expansion work is set to zero in Eq. \eqref{eq:delta_e_I_1}.

The physical process simulated with the barrier method can be thought as an infinitely slow expansion during which some external energy source exactly compensate for the expansion work $\delta W$ against the barrier, or, alternatively, as a series of instantaneous, small, displacements of the barrier, where, after each displacement, one waits for a new equilibrium configuration to establish. 

\subsection{Drift-diffusion approximation}
\label{sec:drift_diff} 

As an alternative to solving Eq. \eqref{eq:Boltzmann}, one can consider the drift-diffusion equation
\begin{equation}
	\dfrac{\partial \Psi}{\partial t} = e\dfrac{\partial \Phi}{\partial t}\dfrac{\partial \Psi}{\partial \epsilon} + \dfrac{2}{3m\nu}\frac{1}{r^2}\dfrac{\partial}{\partial r}\left[r^2\left(\left(\epsilon+e\Phi\right)\dfrac{\partial\Psi}{\partial r}-\frac{e}{2}\dfrac{\partial \Phi}{\partial r}\Psi\right)\right]\text{,}
\label{eq:dPsi_dt}
\end{equation}
obtained from Eq. \eqref{eq:Boltzmann} by approximating $f_e$ as $f_0 (r,v)+{\bf v \cdot \bf f}   _1 (r,v)$ \cite{Raizer}. In Eq. \eqref{eq:dPsi_dt}, the quantity $\Psi\left(r,\epsilon,t\right)$ represents the space-energy distribution of the electrons (i.e., $\Psi\left(r,\epsilon,t\right)\Delta\epsilon$ is the particle density for electrons with energy in the range $\left[\epsilon,\epsilon+\Delta \epsilon\right]$).
The self-consistent potential $\Phi$ is determined by solving Poisson's equation
\begin{equation}
	\frac{1}{r^2}\dfrac{\partial}{\partial r}\left(r^2\dfrac{\partial \Phi}{\partial r}\right) = 4\pi e\left(\int \Psi \diff{\epsilon} - Zn\ped{i0}\right) \text{,}
\label{eq:Poisson_Psi}
\end{equation}
Asymptotically, for $t\rightarrow\infty$, the solution approaches a stationary solution of Eq. \eqref{eq:dPsi_dt}, $\Psi_\infty$, such that 
\begin{equation}
\frac{1}{2}e\dfrac{\diff{\Phi_\infty}}{\diff{r}}\Psi_\infty -\left(\epsilon+e\Phi_\infty\right)\dfrac{\partial\Psi_\infty}{\partial r} = 0\text{.}
\label{eq:current}
\end{equation} 
By solving Eq. \eqref{eq:current} with respect to $\Psi_\infty$, one finds 
\begin{equation}
\Psi_\infty(r,\epsilon) = \dfrac{\rho_\infty(\epsilon)\left[\epsilon+e\Phi_\infty(r)\right]^{\frac{1}{2}}}{4\pi \displaystyle \int_{\mathscr{D}\left(\epsilon\right)} {r^{\prime}}^2\left[ \epsilon+e\Phi_\infty\left(r^{\prime}\right)\right]^{\frac{1}{2}}\diff{r^{\prime}}} \text{,}
\label{eq:Psi_erg}
\end{equation} 
($\rho_\infty(\epsilon)= 4\pi\int\Psi_\infty(r,\epsilon) r^2\diff{r}$ is the energy distribution), which corresponds to the SPE distribution expressed by Eqs. \eqref{eq:rho_e}-\eqref{eq:pr_e}.

\subsection{Results and comparison between models}
\label{sec:results1} 
Examples of initial equilibrium are now presented and discussed, first referring to the two cases of Fig. \ref{fig:ch_tr_all}, then examining the full $\widehat{T}_0$-dependence of the principal equilibrium parameters. A comparison of the self-consistent equilibrium configuration of the electrons after the initial charging transient is made between the exact VP model, Eq. \eqref{eq:VP_1} , the barrier method, and the drift-diffusion model [Eqs. \eqref{eq:dPsi_dt}-\eqref{eq:Poisson_Psi}].

\begin{figure}[!htb]
\centering \epsfig{file=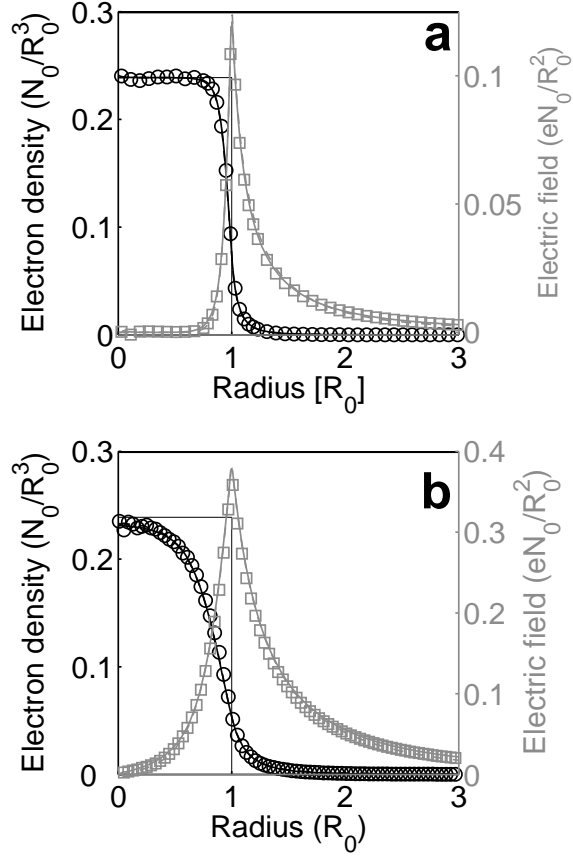, width=3in}
\caption{Equilibrium electron density (black, left axis) and radial electric field (gray, right axis), for (a) $\widehat{T}_0 = 7.2\times 10^{-3}$ and (b) $\widehat{T}_0 = 7.2\times 10^{-2}$. Solid lines refer to results from the ergodic model, markers to results from the full VP model, and dotted lines to results from the drift-diffusion model of Eqs. \eqref{eq:dPsi_dt} and \eqref{eq:Poisson_Psi} (in the plots, the curves obtained with the SPE model and those obtained with the drift-diffusion model are undistinguishable without magnification). Units are normalized to the quantities indicated in parentheses.}
\label{fig:n_E_all}
\end{figure}

\begin{figure}[!htb]
\centering \epsfig{file=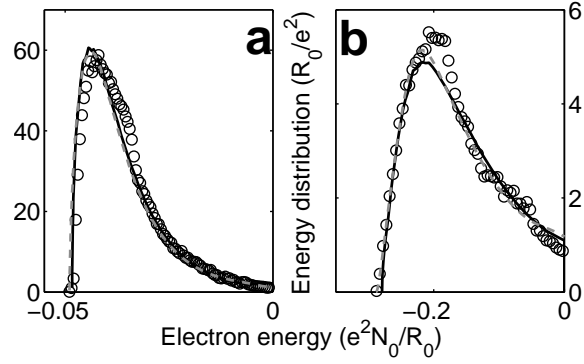, width=3in}
\caption{Equilibrium energy spectrum of trapped ($\epsilon = m{\bf v}^2/2-e\Phi < 0$) electrons, for (a) $\widehat{T}_0 = 7.2\times 10^{-3}$ and (b) $\widehat{T}_0 = 7.2\times 10^{-2}$. Solid lines refer to results from the ergodic model, markers to results from the full VP model, and dotted lines to results from the drift-diffusion model. Units are normalized to the quantities indicated in parentheses.}
\label{fig:rho_E_all}
\end{figure}

\begin{figure}[!htb]
\centering \epsfig{file=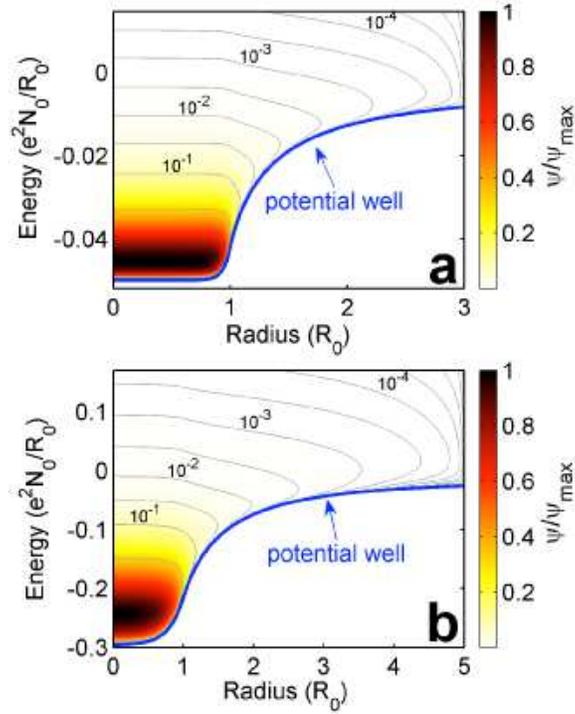, width=3in}
\caption{(Color online) Equilibrium distribution in the $r-\epsilon$ phase space, $\Psi$, as obtained with the drift-diffusion model of Eqs. \eqref{eq:dPsi_dt} and \eqref{eq:Poisson_Psi}, for (a) $\widehat{T}_0 = 7.2\times 10^{-3}$ and (b) $\widehat{T}_0 = 7.2\times 10^{-2}$. The spatial-energetic distribution is normalized to its maximum value $\Psi\ped{max}$ and isolevel curves are plotted at $10^{-n/2}\Psi\ped{max}$ (where $n$ is an integer). Units are normalized to the quantities indicated in parentheses.}
\label{fig:psi_all}
\end{figure}

In Fig. \ref{fig:n_E_all}, the electron density is plotted, along with the corresponding electric field: the positive charge buildup at the ion front, $\Delta Q$, is (a) $12.5\%$ and (b) $38\%$ of the total ionic charge $eN_0$. Figure \ref{fig:rho_E_all} shows the equilibrium energy distribution $\rho\ped{e,0}$, to be used as initial condition for the bulk expansion. The excellent agreement between different models confirms the validity of the barrier method. Figure \ref{fig:psi_all} shows the asymptotic solution of Eqs. \eqref{eq:dPsi_dt} and \eqref{eq:Poisson_Psi}, $\Psi_\infty(r,\epsilon)$: the corresponding electron density and energy distribution, plotted in Figs. \ref{fig:n_E_all} and \ref{fig:rho_E_all}, have been calculated as $\int\Psi_\infty(r,\epsilon)\diff{\epsilon}$ and $4\pi\int\Psi_\infty(r,\epsilon)r^2\diff{r}$, respectively. 

\begin{figure}[!htb]
\centering \epsfig{file=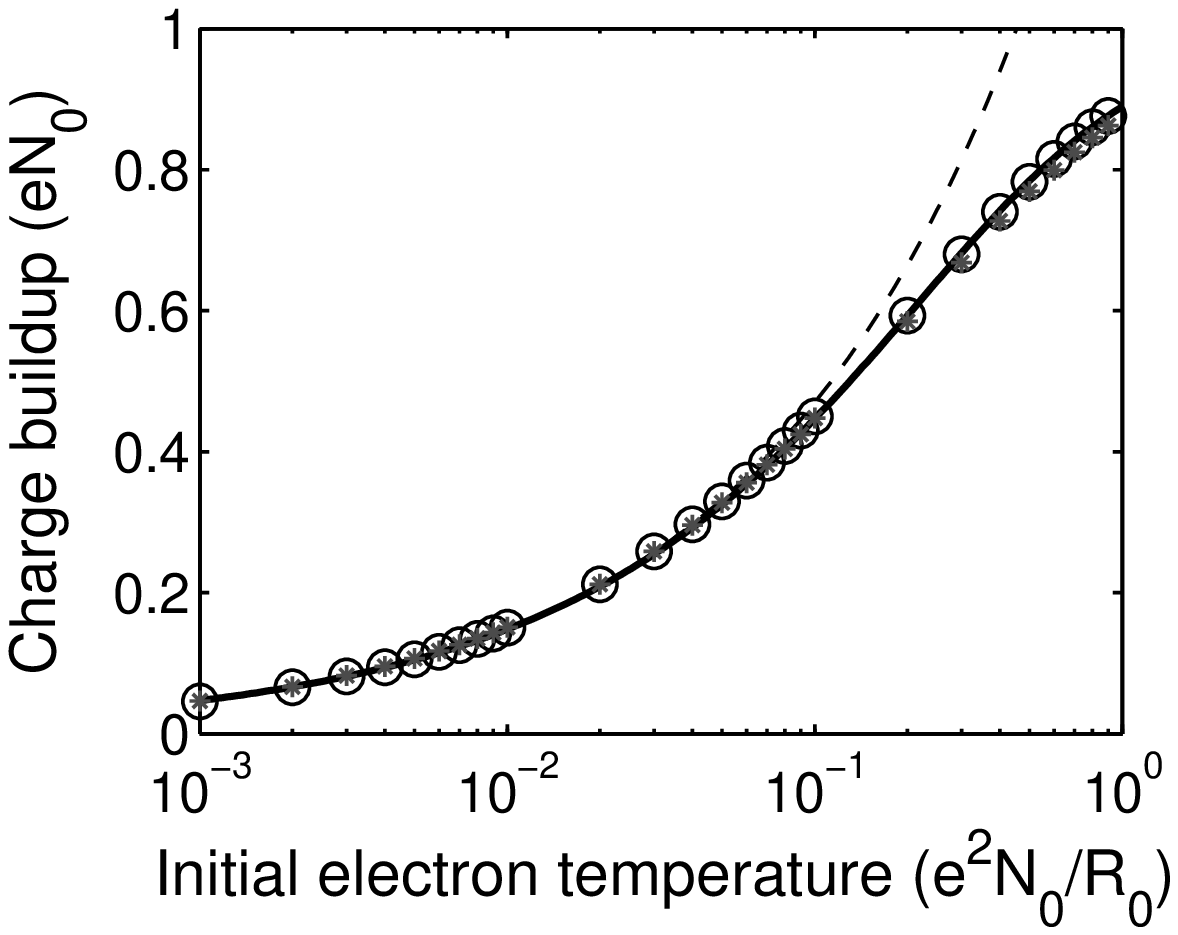, width=2.5in}
\caption{Equilibrium charge buildup as a function of $\widehat{T}_0$. Markers are used for results obtained with the SPE model (circles) and the drift-diffusion model (asterisks); the solid line represent the fit law of Eq. \eqref{eq:fit1}, whereas the dotted line shows the corresponding power-law behavior for $\widehat{T}_0 \ll 1$. Units are normalized to the quantities indicated in parentheses.}
\label{fig:Qeq}
\end{figure}

\begin{figure}[!htb]
\centering \epsfig{file=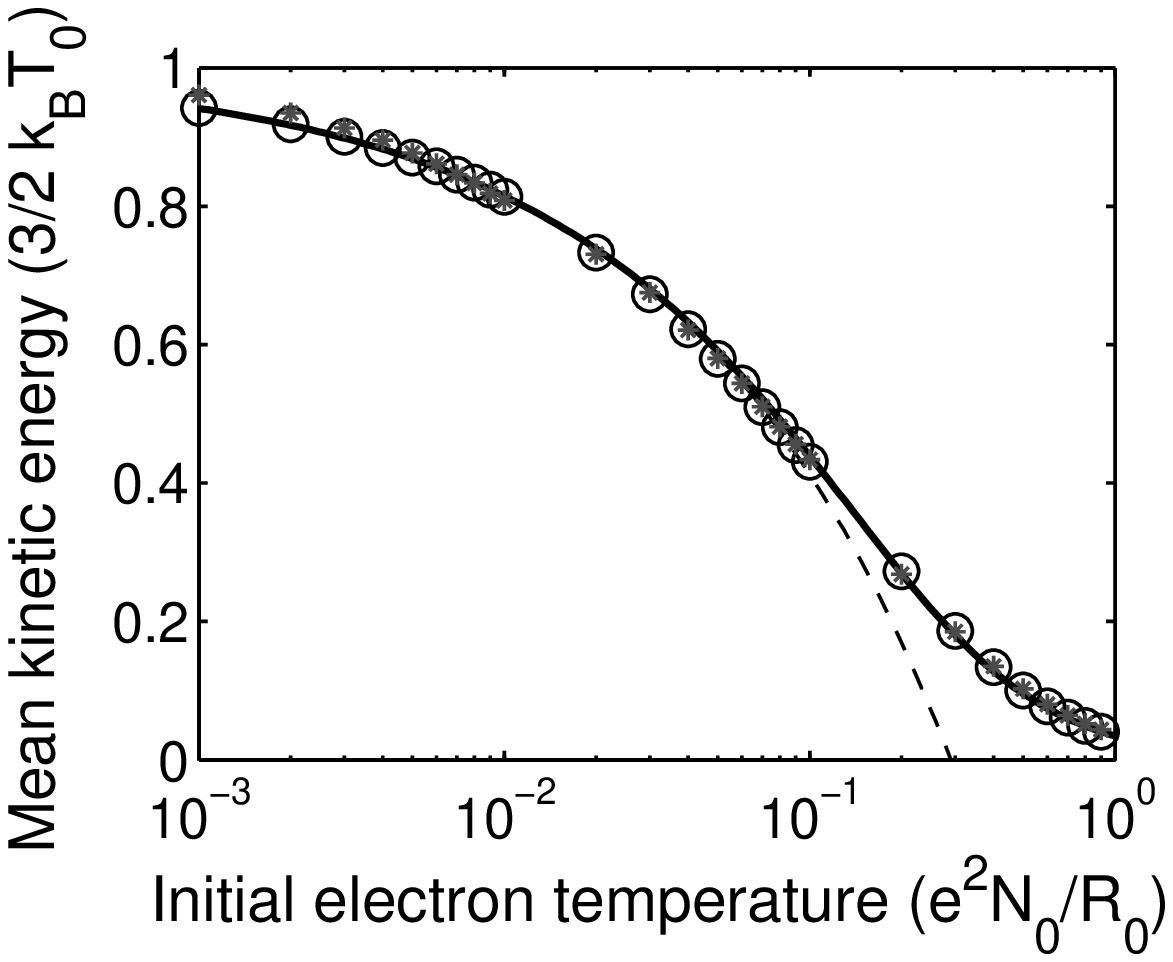, width=2.5in}
\caption{Equilibrium total kinetic energyas a function of $\widehat{T}_0$. Markers are used for results obtained with the SPE model (circles) and the drift-diffusion model (asterisks); the solid line represent the fit law of Eq. \eqref{eq:fit2}, whereas the dotted line shows the corresponding power-law behavior for $\widehat{T}_0 \ll 1$. Units are normalized to the quantities indicated in parentheses.}
\label{fig:Teq}
\end{figure}

The dependence of the initial equilibrium on $\widehat{T}_0$ has been analyzed using both the barrier method and the drift-diffusion approximation, for $\widehat{T}_0$ varying in the range $[10^{-3}, 1]$: the equilibrium values of $\Delta Q$ and of the mean kinetic energy of the trapped electrons, $\mathcal{E}$, are displayed in Figs. \ref{fig:Qeq} and \ref{fig:Teq}, respectively, along with the corresponding fit laws  (obtained in Ref. \cite{Peano_PRL_2} using the SPE model),
\begin{equation}
	\frac{\Delta Q}{eN_0} = \mathcal{F}_{2.60}\left(\sqrt{6/e}\widehat{T}_0^{1/2}\right)
	\text{,}
\label{eq:fit1}
\end{equation}
\begin{equation}
	\frac{\mathcal{E}}{\frac{3}{2}k\ped{B}T_0}=
	1-\mathcal{F}_{3.35}\left(1.86\widehat{T}_0^{1/2}\right)
	\text{,}
\label{eq:fit2}
\end{equation}
where $\mathcal{F}_{\mu}(x) = x/(1+x^{\mu})^{1/\mu}$, and where the coefficient $\sqrt{6/e}$ in Eq. \eqref{eq:fit1} provides a match with the analytical results for the planar case \cite{Crow,Mora} in the limit $\widehat{T}_0 \ll 1$. Again, an excellent agreement is found between different calculations.

\section{Bulk expansion}
The self-consistent expansion of ions and electrons has been investigated for a wide range of the parameter $\widehat{T}_0$ by solving Eqs. \eqref{eq:model}, having used the barrier method to determine the initial equilibrium distribution of electrons. The results of the study reveal that the expansion dynamics changes smoothly from a hydrodynamic-like regime (in which the outer ions expand first and a rarefaction front propagates inward) to a CE-like regime (in which all ions start expanding at the same time), when going from $\widehat{T}_0 \ll 1$ to $\widehat{T}_0 \sim 1$. 
Nonetheless, a qualitative change in the ion energy spectrum is observed for $\widehat{T}_0 \simeq 0.5$, marking the transition towards a CE behavior. 
Following the organization of Sec. \ref{sec:results1}, the bulk expansion is first analyzed in detail for the two reference cases [cases (a) and (b) henceforth], in which (a) $\widehat{T}_0=7.2\times10^{-3}$ and (b) $\widehat{T}_0=7.2\times10^{-2}$, and then the dependence of the most relevant expansion features on $\widehat{T}_0$ is examined.

\begin{figure}[!htb]
\centering \epsfig{file=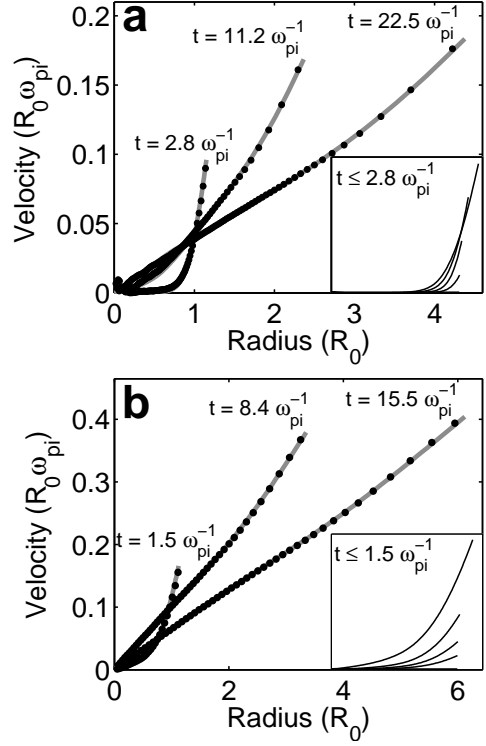, width=2.5in}
\caption{Evolution of the ion phase-space profile, for (a) $\widehat{T}_0 = 7.2\times 10^{-3}$ and (b) $\widehat{T}_0 = 7.2\times 10^{-2}$. Lines refer to results from the ergodic model, markers to results from the full VP model. Insets show the evolution of the ion phase-space profile during the early stage of the expansion. Units are normalized to the quantities indicated in parentheses.}
\label{fig:phsp}
\end{figure}

\begin{figure}[!htb]
\centering \epsfig{file=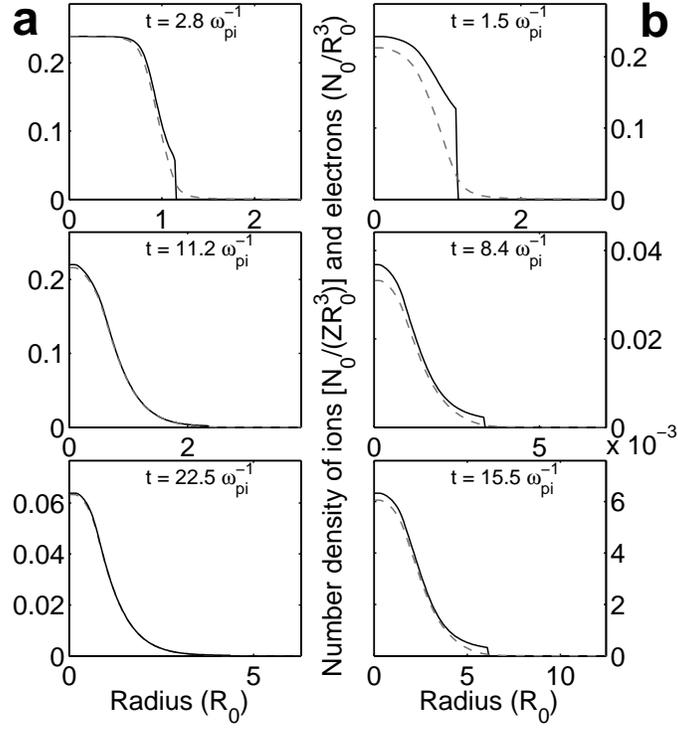, width=3.5in}
\caption{Evolution of ion (solid line) and electron (dashed curve) density, for (case a, left panels) $\widehat{T}_0 = 7.2\times 10^{-3}$ and (case b, right panels) $\widehat{T}_0 = 7.2\times 10^{-2}$. Units are normalized to the quantities indicated in parentheses.}
\label{fig:densities}
\end{figure}

\begin{figure}[!htb]
\centering \epsfig{file=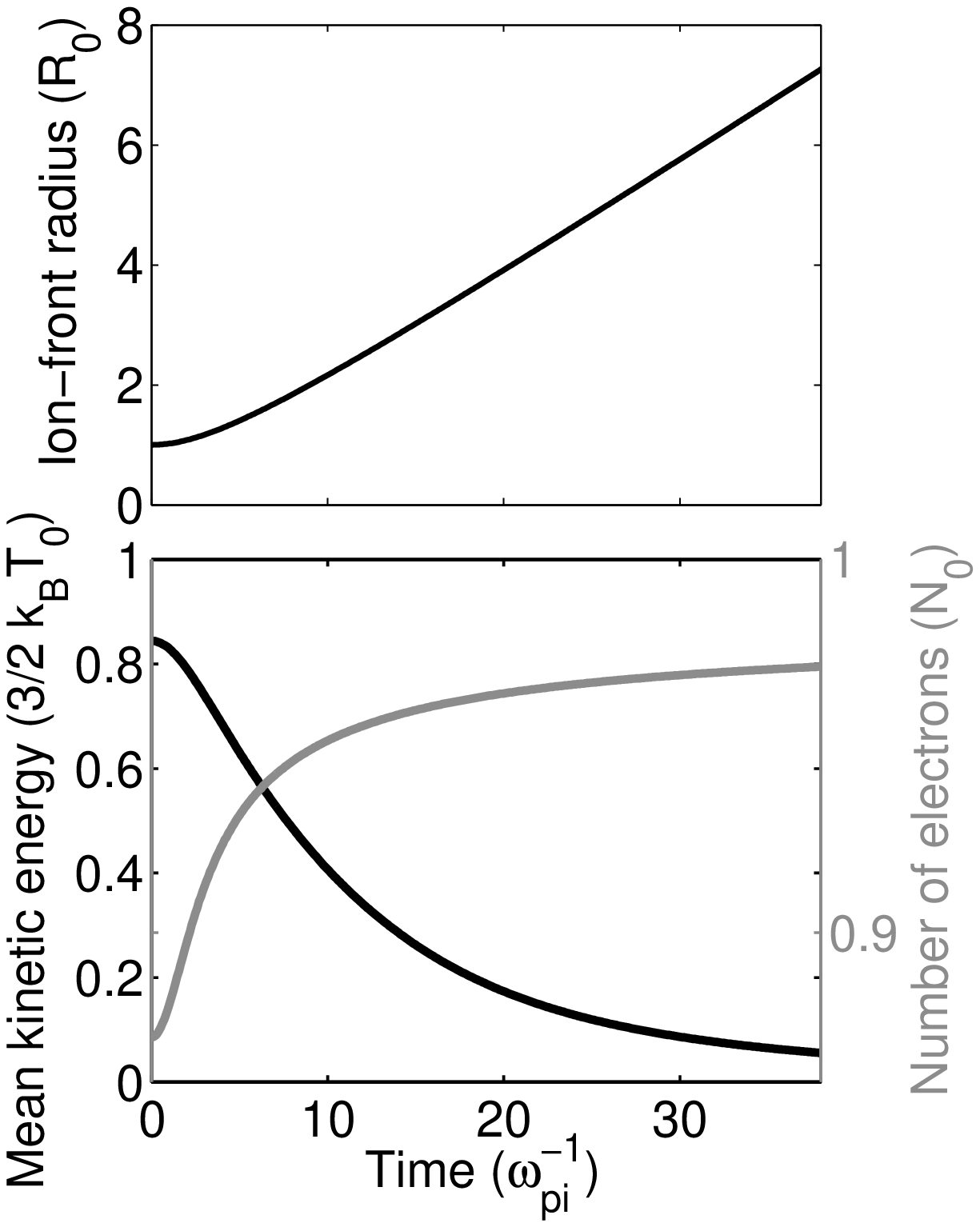, width=2.5in}
\caption{Radial trajectory of the ion-front (top panel), evolution of the number of electrons enclosed by the ion front (bottom panel, gray), and evolution of the mean kinetic energy of trapped electrons (bottom panel, black), for $\widehat{T}_0= 7.2\times 10^{-3}$. Units are normalized to the quantities indicated in parentheses.}
\label{fig:Q_T_time_a}
\end{figure}

\begin{figure}[!htb]
\centering \epsfig{file=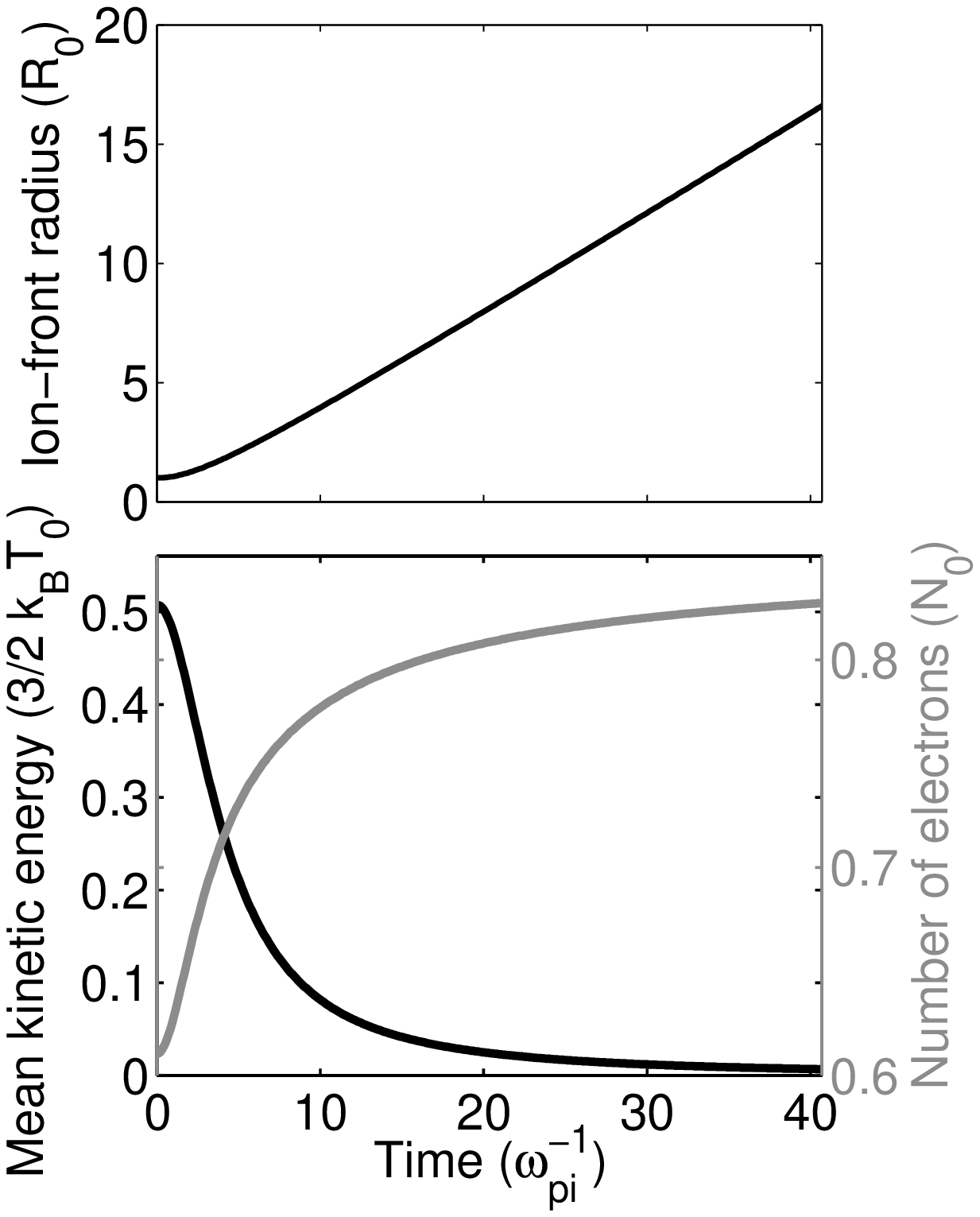, width=2.5in}
\caption{Radial trajectory of the ion-front (top panel), evolution of the number of electrons enclosed by the ion front (bottom panel, gray), and evolution of the mean kinetic energy of trapped electrons (bottom panel, black), for $\widehat{T}_0= 7.2\times 10^{-2}$. Units are normalized to the quantities indicated in parentheses.}
\label{fig:Q_T_time_b}
\end{figure}

\begin{figure}[!htb]
\centering \epsfig{file=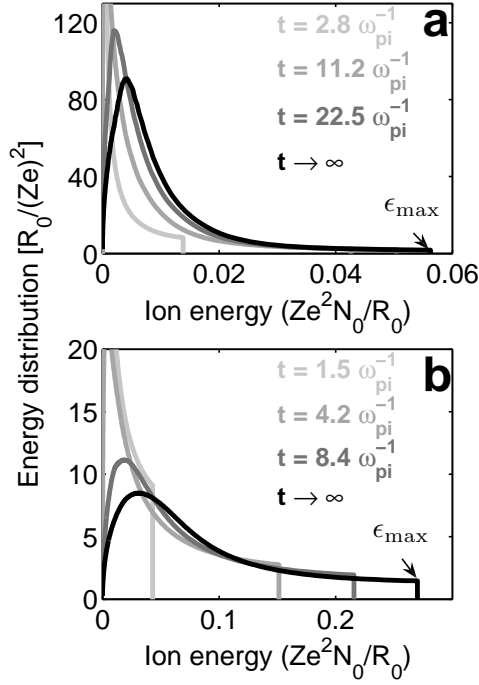, width=2.5in}
\caption{Evolution of the ion energy spectrum (from light gray to black), for (a) $\widehat{T}_0 = 7.2\times 10^{-3}$ and (b) $\widehat{T}_0 = 7.2\times 10^{-2}$. Units are normalized to the quantities indicated in parentheses.}
\label{fig:spectrum_all}
\end{figure}

\begin{figure}[!htb]
\centering \epsfig{file=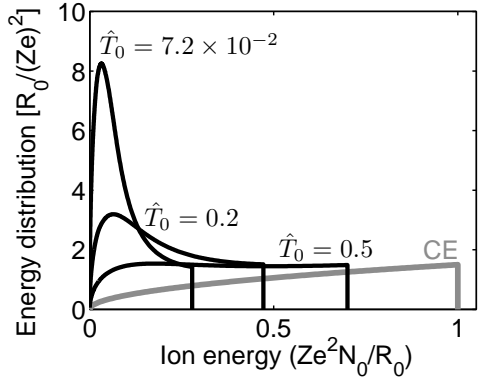, width=2.5in}
\caption{Asymptotic ion energy spectra for different values of $\widehat{T}_{0}$, compared with the theoretical asymptotic spectrum for the CE case (gray curve). Units are normalized to the quantities indicated in parentheses.}
\label{fig:spectrum_T0}
\end{figure}

\begin{figure}[!htb]
\centering \epsfig{file=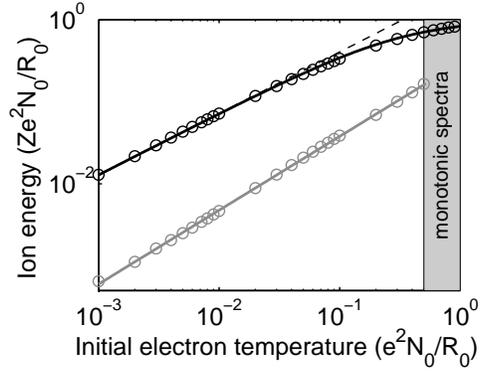, width=2.5in}
\caption{Cutoff ion energy (black) and location of the maximum in the ion energy spectrum (gray) as functions of $\widehat{T}_0$: circles refer to the SPE model, solid lines to the fit laws in the text. The dashed line represents the power-law behavior of $\epsilon\ped{max}$ for $\widehat{T}_0 \ll 1$. Units are normalized to the quantities indicated in parentheses.}
\label{fig:ion_energy}
\end{figure}

The evolution of the ion phase-space profile and of the electron and ion densities (starting from the initial equilibrium of Fig. \ref{fig:n_E_all}), are shown in Figs. \ref{fig:phsp} and \ref{fig:densities}, respectively. In case (a) (Figs. \ref{fig:phsp}a and \ref{fig:densities}a),
the ion expansion starts from the periphery and a rarefaction front is clearly observed to propagate inward until it reaches the center of the distribution; during the expansion, the plasma remains approximately neutral, apart from the thin double-layer at the ion front. These features, typical of quasineutral, hydrodynamic expansions, are lost in case (b),
in which all the ions are promptly involved in the expansion (Fig. \ref{fig:phsp}b) and the distribution remains nonneutral during the whole process (Fig. \ref{fig:densities}b).
In both cases, as the ions expand, and gain kinetic energy, the electrons cool down and the charge buildup within the ion sphere decreases, as illustrated in Fig. \ref{fig:Q_T_time_a} [case (a)] and Fig. \ref{fig:Q_T_time_b} [case (b)]. Asymptotically, the sphere enveloped by the expanding ion front encloses all trapped electrons, and a ballistic regime is reached for both species \cite{Manfredi}.
The self-consistent behavior of the electrons strongly affects the ion dynamics and their resulting energy spectrum.
In fact, starting from the equation of motion of the ions [the first of Eqs. \eqref{eq:model}], the asymptotic energy $\epsilon_\infty$ of an ion can be written as 
\begin{equation}
\frac{\epsilon_\infty(r_0)}{Ze} = \frac{q(r_0,0)}{r_0} + \int^{\infty}_{0} \! \!  \frac{1}{r\ped{i}(r_0,t)}\frac{\partial q\left(r\ped{i}(r_0,t),t\right)}{\partial t}\diff{t}
\label{eq:E_ion} \text{,}
\end{equation}
where $q(r,t)$ is the net charge buildup enclosed by a sphere of radius $r$ at time $t$. The first term on the right-hand side of Eq. \eqref{eq:E_ion} is the ion potential energy, whereas the integral term (vanishing for a CE) accounts for the energy loss due to the decreasing charge buildup experienced by the ions along their trajectory. Figure \ref{fig:spectrum_all} illustrates the evolution of the ion energy spectrum towards its asymptotic form: in both cases, the spectrum develops a well-defined local maximum far from the cutoff energy. 
Since this feature is absent in CEs, where the asymptotic spectrum is always monotonic (it behaves as $\epsilon^{1/2}$ up to the cutoff energy $\epsilon\ped{CE}$), the maximum in the spectrum is expected to disappear when increasing $\widehat{T}_0$ further. This transition from nonmonotonic to monotonic ion spectra occurs about $\widehat{T}_0 = 0.5$ (cf. Fig. \ref{fig:spectrum_T0}) and marks the transition towards a CE-like behavior. In this sense, $\widehat{T}_0 = 0.5$ can be considered as a lower bound for the validity of the CE model. The dependence of the maximum (cutoff) ion energy $\epsilon\ped{max}$ on $\widehat{T}_0$ is shown in Fig. \ref{fig:ion_energy}, along with the energy value of the local maximum in the spectrum, $\epsilon\ped{peak}$. The behavior of $\epsilon\ped{max}$ is accurately described by the fit law
\begin{equation}
\epsilon\ped{max}=\mathcal{F}_{1.43}\left(2.28\:\widehat{T}_0^{3/4}\right)\epsilon\ped{CE}\text{,}
\label{eq:fit3}
\end{equation}
($\mathcal{F}$ belongs to the same class of functions used in Eqs. \eqref{eq:fit1} and \eqref{eq:fit2}), whereas $\epsilon\ped{peak}$ exhibits the power-law behavior $\epsilon\ped{peak}=0.3\widehat{T}_0^{0.9}\epsilon\ped{CE}$, for $\widehat{T}_0 < 0.5$.
These fit laws can be used to provide useful estimates of the initial electron temperature and, hence, of the expansion regime. This can be important for the interpretation of experimental ion-spectrum data. In fact, for expansion conditions far from a CE, the nonmonotonic behavior of the single-cluster ion spectra could affect the total (i.e., arising from all expanding clusters) energy spectrum measured in experiments: in particular, it could lead to nonmonotonic energy spectra, such as those presented in Ref. \cite{Sakabe}, also for narrow distributions of cluster radii. 

\section{Conclusions}
The results presented in the paper prove that the collisionless expansion of spherical plasmas driven by hot electrons can be analyzed accurately with a kinetic model that describes the electron distribution as a sequence of ergodic equilibrium configurations. The self-consistent equilibrium that is established after the initial, sudden expansion of the electrons has been investigated in detail, in the frozen-ion approximation. 
This equilibrium can be determined with great accuracy by replacing the real (fast) transient with an appropriate virtual (slow) process, finding excellent agreement with reference solutions of the full VP model. 
This guarantees a highly-precise description of the whole process, thus providing an effective tool for the analysis of the expansion dynamics. In particular, a transition in the behavior of the ion energy spectrum, when approaching the Coulomb-explosion regime, has been identified, and accurate fit laws for the general properties of the expansion, which are valid for any value of dimensionless electron temperature  (provided that relativistic effects are negligible), have been determined. 
These laws can furnish useful estimates for the interpretation of experimental data, in particular concerning possible influences of single-cluster effects on measured ion spectra.

Finally, the ergodic model presented here can be readily employed to study more general physical situations, such as expansions driven by initially non-Maxwellian electrons. Furthermore, the model could be extended so as to include relativistic velocities and to account for the effects of non-instantaneous electron heating by ultraintense laser pulses. 

\begin{acknowledgments}
Work partially supported by FCT (Portugal) through Grant No. POCI/FIS/55095.
\end{acknowledgments}

\newpage
\section{Figures}

\end{document}